\documentclass[10pt,aps,prl,twocolumn,superscriptaddress,floatfix,amssymb,showpacs]{revtex4-1}
\usepackage{amsmath,amssymb,amsfonts,color}
\usepackage{graphicx}
\usepackage{subfigure}
\usepackage{mathbbol,bbm,bm}

\begin{document}

\title{Nature of the Spin Liquid Ground State of the $S=1/2$ Heisenberg Model on the Kagome Lattice}

\author{Stefan Depenbrock}
\email{stefan.depenbrock@lmu.de}

\affiliation{Department of Physics and Arnold Sommerfeld Center for Theoretical Physics,
Ludwig-Maximilians-Universit\"at M\"unchen, 80333 M\"unchen, Germany}
\author{Ian P. McCulloch}
\affiliation{Centre for Engineered Quantum Systems,
School of Mathematics and Physics,
The University of Queensland, St Lucia QLD 4072, Australia}

\author{Ulrich Schollw{\"o}ck}
\affiliation{Department of Physics and Arnold Sommerfeld Center for Theoretical Physics,
Ludwig-Maximilians-Universit\"at M\"unchen, 80333 M\"unchen, Germany}

\date{\today}

\begin{abstract}
We perform a density-matrix renormalization group (DMRG) study of the $S=\frac{1}{2}$ Heisenberg antiferromagnet on the kagome lattice to identify the conjectured spin liquid ground state. Exploiting SU(2) spin symmetry, which allows us to keep up to 16~000 DMRG states, we consider cylinders with circumferences up to 17 lattice spacings and find a spin liquid ground state with an estimated per site energy of $-0.4386(5)$, a spin gap of $0.13(1)$, very short-range decay in spin, dimer and chiral correlation functions and finite topological entanglement $\gamma$ consistent with $\gamma=\log_2 2$, ruling out gapless, chiral or non-topological spin liquids in favor of a topological spin liquid of quantum dimension 2, with strong evidence for a gapped topological $\mathbb{Z}_2$ spin liquid.
\end{abstract}


\maketitle

A pervasive feature of physics is the presence of symmetries and their breaking at low energies and temperatures. It would be an unusual system in which at $T=0$ (quantum) fluctuations are so strong that all symmetries remain unbroken in the ground state. In magnetic systems such a state is dubbed a quantum spin liquid (QSL)\cite{Balents_2010} and is most likely to occur if fluctuations are maximized by low-dimension, low-spin and strong geometrical frustration; the search for a QSL has thus focused on frustrated $S=\frac{1}{2}$ quantum magnets in two dimensions. The Heisenberg antiferromagnet on the kagome lattice\cite{elser_1989} (KAFM) is a key candidate, described by the $S=\frac{1}{2}$ model 
\begin{equation}
\mathcal{H} = \sum_{\langle i, j \rangle} \vec{S}_i \cdot \vec{S}_j,
\label{eq:kagome}
\end{equation}
with $\langle i, j \rangle$ nearest neighbors. 

Experimentally, the focus is on the Herbertsmithite $\mathrm{ZnCu_3(OH)_6Cl_2}$, modeled by (\ref{eq:kagome}) on a kagome lattice with additional Dzyaloshinskii-Moriya interactions\cite{matan_dzyaloshinskii-moriya_2011}. It is thought that the ground state is a spin liquid\cite{Olariu_2008,Bert_2007,mendels_quantum_2006,Imai_2008,Lee_2007,deVries_2008,mendels_quantum_2011}, with no onsite magnetization \cite{mendels_quantum_2006,Helton_2007} and no spin gap \cite{Helton_2007,imai_local_2011,jeong_field-induced_2011,Wulferding_2010} within very tight experimental bounds. 

On the theoretical side, the kagome model of Eq.~(\ref{eq:kagome}) remains a formidable challenge. While all proposed ground states show no magnetic ordering, they can be classified by whether they break translational invariance or not. The former type of ground state, a valence bond crystal (VBC), was pioneered by Marston\cite{marston_spinpeierls_1991}. The emerging proposal was that of a ``honeycomb VBC'' (HVBC) with a hexagonal unit cell of 36 spins \cite{hastings_dirac_2000, Nikolic_2003,singh_ground_2007,singh_triplet_2008,iqbal_valence-bond_2012} sharing in dimer-covered hexagons and a sixfold ``pin wheel'' at the center. On the other hand, a multitude of QSL states were proposed\cite{messio_kagome_2011, Yang_QSL, hermele_properties_2008, iqbal_projected_2011, ran_projected_2007, ryu_algebraic_2007, sachdev_kagome-_1992, wang_sbmft, lu_z_2_2011, lu_z_2_2011, Misguich_dimer_2002, jiang_density_2008, huh_vison_2011}.  Proposals for a QSL ground state include a chiral topological spin liquid\cite{Kalmeyer_1989,wen_chiral,Yang_QSL,messio_kagome_2011}, a gapless spin liquid\cite{hermele_properties_2008, iqbal_projected_2011,ran_projected_2007, ryu_algebraic_2007}, and various $\mathbb{Z}_2$ spin liquids\cite{sachdev_kagome-_1992, wang_sbmft, lu_z_2_2011, Misguich_dimer_2002} with topological ground state degeneracy.  

In the past, numerical methods failed to resolve the issue conclusively. Quantum Monte Carlo faces the sign problem. Sizes accessible by exact diagonalization \cite{elser_1989, zeng_1990, chalker_1992, leung_1993, elstner_1994, lecheminant_1997, waldtmann_chiral, sindzingre_2000, waldtmann_2000, richter_2004, laeuchli_dynamical_2009,  sindzingre_low_2009, sorensen_2009, lauchli_ground-state_2011,Nakano_2011} are currently limited to 48 sites. Other approaches diagonalized the valence bond basis or applied the contractor renormalization group (CORE) method, or the coupled cluster method (CCM) \cite{zeng_quantum_1995, mambrini_2000, budnik_2004, capponi_2004, poilblanc_2010, schwandt_generalized_2010,Richter_2011}.
The multi-scale entanglement renormalization ansatz (MERA) \cite{evenbly_frustrated_2010} found the VBC state lower in energy than the QSL state reported in an earlier density-matrix renormalization group (DMRG) study of tori up to 120 sites \cite{jiang_density_2008}. 

Recently, strong evidence for a QSL was found in a large-scale DMRG study \cite{yan_spin_2011}  considering long cylinders of circumference up to 12 lattice spacings. Ground state energies were substantially lower than those of the VBC state and an upper energy bound substantially below the VBC state energy was found; the ground state, having the hallmarks of a QSL, was not susceptible to attempts to enforce a VBC state. As to the type of QSL, \cite{yan_spin_2011} did not provide direct evidence for a $\mathbb{Z}_2$ topological QSL. This has sparked a series of papers trying to identify the QSL\cite{huh_vison_2011,iqbal_projected_2011, poilblanc_competing_2011, schwandt_generalized_2010, messio_kagome_2011}, where again chiral spin liquids and gapless U(1) spin liquids were advocated and a classification of  $\mathbb{Z}_2$ spin liquids achieved. At the moment, the issue is not conclusive.

Here we study the KAFM using DMRG \cite{white_density_1992, schollwock_density-matrix_2005,ulrich_density-matrix_2011}, in the spirit of \cite{yan_spin_2011}. DMRG is a variational method in the ansatz space spanned by matrix product states (MPS) which allows it to find the ground state of one-dimensional (1D) systems efficiently even for large system sizes. It can also be applied successfully to two-dimensional (2D) lattices by mapping the short-ranged 2D Hamiltonian exactly to a long-ranged 1D Hamiltonian \cite{White_spin-gaps_1996, White_tj_1998, White_energetics_1998, white_neel_2007, yan_spin_2011, stoudenmire_studying_2011}. DMRG cost scales roughly exponentially with entanglement entropy, such that area laws limit system sizes, and DMRG favors open boundary conditions (OBCs) over preferable periodic boundary conditions (PBCs). The conventional compromise \cite{yan_spin_2011}, taken also by us, is to consider cylinders, i.e.\ PBCs along the short direction (circumference $c$) and OBCs along the long direction (length $L$) where boundary effects are less important. Cost is dominated exponentially by circumference $c$. We use two different 1D mappings (labeled as XC and YC plus cylinder size) \cite{SM} to check for undesired mapping dependencies of the DMRG results. Instead of earlier Abelian U(1) DMRG with up to 8~000 ansatz states, we employ non-Abelian SU(2) DMRG \cite{mcculloch_density-matrix_2007,mcculloch_non-abelian_2000} based on irreducible representations corresponding to 16~000 ansatz states in a U(1) approach. This has crucial advantages: available results can be verified with much higher accuracy. The circumference of the cylinders can be increased by almost 50 \% from 12 to 17.3 lattice sites (up to 726 sites in total), strongly reducing finite size effects; we also consider tori of up to 108 sites. We can eliminate the spin degeneracy that necessitates pinning fields in U(1)-symmetric simulations and avoid artificial constraints in gap calculations, making them more accurate and reliable. We also present results on spin, dimer and chiral correlation functions, the structure factor and topological entanglement entropy. All data agree with a gapped non-chiral $\mathbb{Z}_2$ spin liquid; other QSL proposals for the KAFM are inconsistent with at least one of the numerical results.   

\begin{figure}
  \includegraphics[height=160pt]{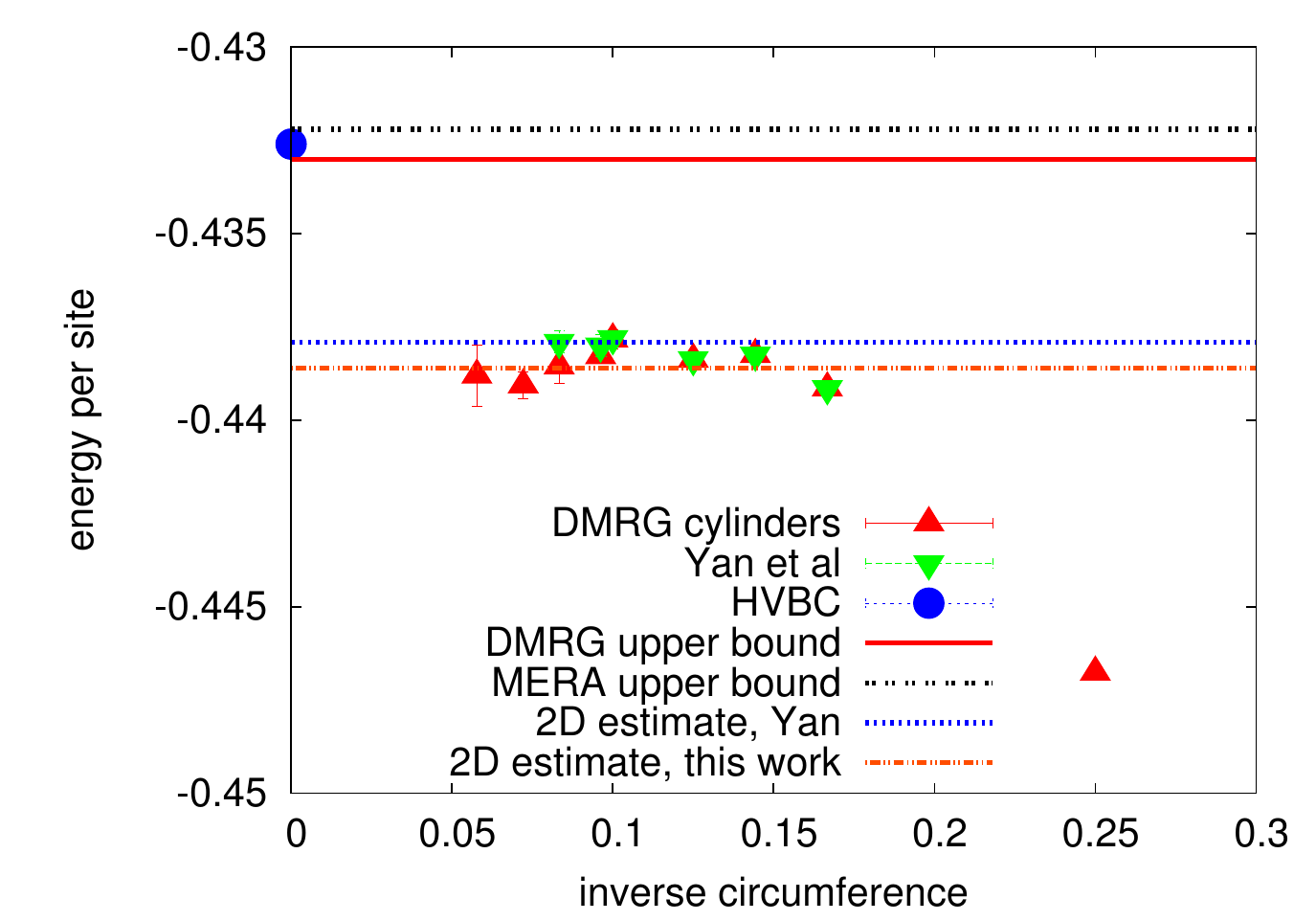}
  \caption{\label{fig:energy}Bulk energies per site. Lengths are in units of lattice spacings. The HVBC result \cite{singh_ground_2007, singh_triplet_2008}, and the upper bounds of
MERA \cite{evenbly_frustrated_2010} and DMRG \cite{yan_spin_2011} apply
directly to the thermodynamic limit; 2D estimates are extrapolations.}
\end{figure}

{\em Energies.}--Energies for cylinders of fixed $c$ and $L$ are extrapolated in the truncation error of single-site DMRG \cite{White_single_2005}; bulk energies per site are extracted by a subtraction technique \cite{stoudenmire_studying_2011} and extrapolated to $L\rightarrow\infty$. Results for various 1D mappings and $c$ are displayed in Table \ref{tab:energy}. We also show the spin (triplet) gap to the $S=1$ spin sector.
\begin{table}
% \begin{ruledtabular}
\begin{tabular}{|l|l|l|l|l|l|}
& $c$ &  $E/N$ & gap $\Delta_E$ & $E_{\mathrm{earlier}}$ & $\Delta_{E,\mathrm{earlier}}$ \\ \hline
YC4   & $4$ & $-0.446~77$ & $0.2189$ & $-0.4467$ & \\
YC6   & $6$ & $-0.439~15(5)$ & $0.1396(6)$ & $-0.439~14$ & $0.142(1)$ \\
YC8   & $8$ & $-0.438~38(5)$ & $0.135(3)$ & $-0.438~36(2)$ & $0.156(2)$ \\
YC10  & $10$ & $-0.4378(2)$ & & $-0.4378(2)$ & $0.070(15)$ \\
YC12  & $12$ & $-0.4386(4)$ & & $-0.4379(3)$ & \\
XC8   & $6.9$ &  $-0.43826(4)$ & $0.13899(1)$ & $-0.438~24(2)$ & $0.1540(6)$  \\
XC12  & $10.4$ & $-0.438~29(7)$ & $0.134(4)$ & $-0.4380(3)$ & $0.125(9)$ \\
XC16  & $13.9$ & $-0.4391(3)$ & $0.130(7)$ &  &  \\
XC20  & $17.3$ & $-0.4388(8)$ & & &  \\
Torus & 3 &$-0.436~278$ & $0.2687$ & $-0.436~278$ & $0.2687$ \cite{lauchli_ground-state_2011} \\ 
Torus & 4 &$-0.4383(2)$ & $0.151$ & $-0.435~91$ & $0.140$ \cite{jiang_density_2008} \\
Torus & 6 &$-0.4383(3)$ & $0.1148(1)$ & $-0.43111$ & $0.105$ \cite{jiang_density_2008}
\end{tabular}
% \end{ruledtabular}
\caption{\label{tab:energy} Ground state energy per site ($E/N$) and gaps for $L=\infty$ cylinders (circumference $c$). Errors are from extrapolation; comparisons are with Ref.~\cite{yan_spin_2011} except for the tori.}
\end{table}
We confirm and extend earlier results \cite{yan_spin_2011}. At 16~000 states, DMRG is highly accurate; negligible changes in energy for substantially larger $c$ support that the thermodynamic limit energy is found, which we place at $-0.4386(5)$ (Fig. \ref{fig:energy}). Similar to Ref.~\cite{yan_spin_2011} we find the energy to be significantly below that of VBC states and no trace of a VBC in the correlation patterns. Except for the edges, bond energies are fully translationally invariant. All results are consistent with strict variational upper bounds obtained without extrapolations from independent DMRG calculations for infinitely long cylinders using the iDMRG variant \cite{iDMRG}, which are below the VBC energies. 

\begin{figure}[hb]
  \includegraphics[width=0.5\textwidth]{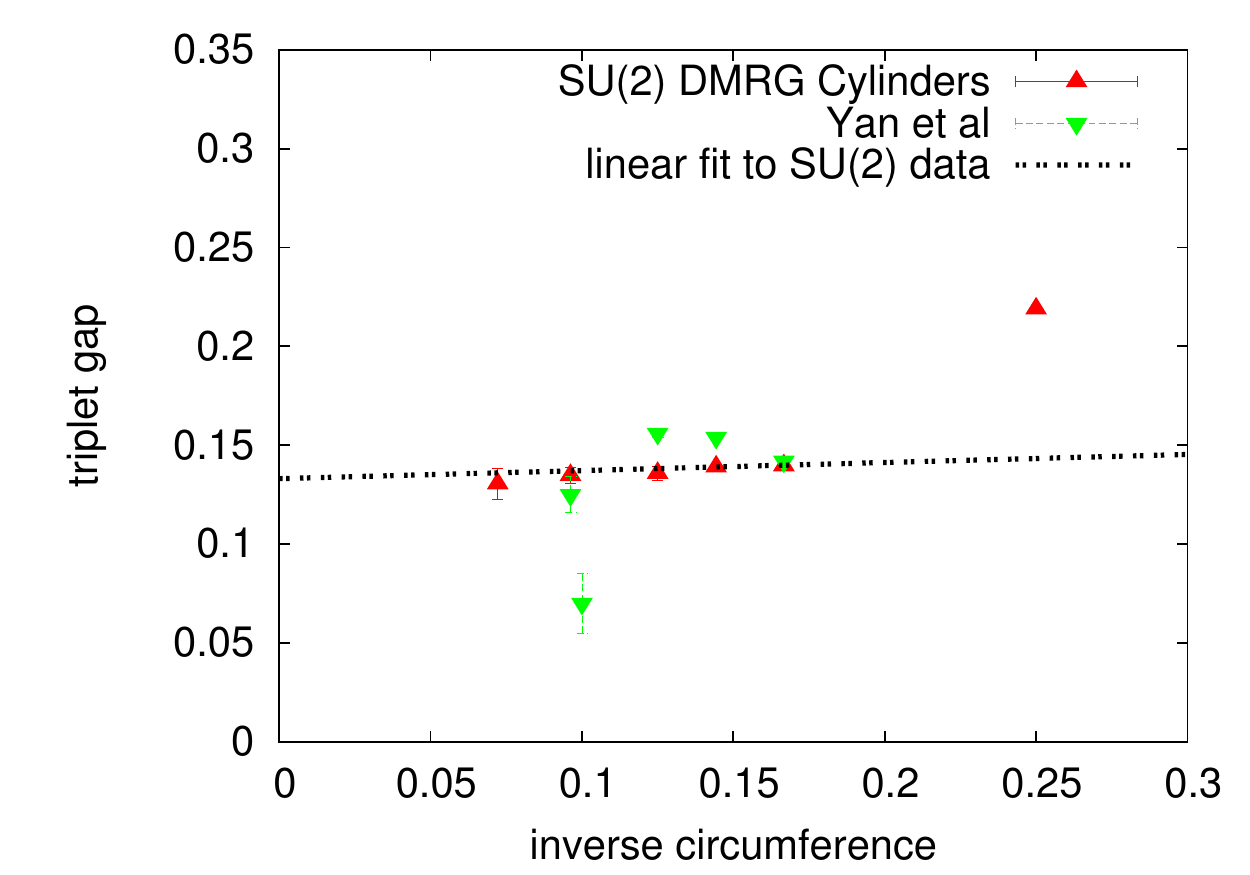}
  \caption{\label{fig:gap}Plot of the bulk triplet gap for infinitely long cylinders versus the inverse
circumference $c$ in units of inverse lattice spacings with an empirical linear fit to the largest cylinders, leading to a spin gap estimate of $0.13(1)$.}
\end{figure}

On the issue of a spin (triplet) gap \cite{sindzingre_quantum_2001,sindzingre_low_2009}, Yan \textit{et al.} \cite{yan_spin_2011} argue in favor of a small, but finite spin gap. SU(2) DMRG computes the $S=1$ state directly and more efficiently; boundary excitations are excluded by examining local bond energies. We find the spin gap (Table \ref{tab:energy} and Fig. \ref{fig:gap}) to remain finite also for cylinders of large $c$. Whereas the results for small $c$ agree with the $S=1$ state energies and gaps reported in \cite{yan_spin_2011}, they display significant differences for larger $c$, perhaps due to the more complex earlier calculation scheme. SU(2)-invariant results evolve more smoothly with $c$, allowing a tentative extrapolation to a spin gap $\Delta_E = 0.13(1)$ in the thermodynamic limit. Size dependence is small, in line with very short correlation lengths. The finite spin gap contradicts conjectures of a $U(1)$ or other gapless spin liquids. For the calculation of the singlet gap found to be finite in Ref.~\cite{yan_spin_2011}, SU(2) DMRG does not offer a significant advantage to be reported here.

\begin{figure}
\subfigure[\label{fig:corrfit}]{
\includegraphics[width=0.4\textwidth]{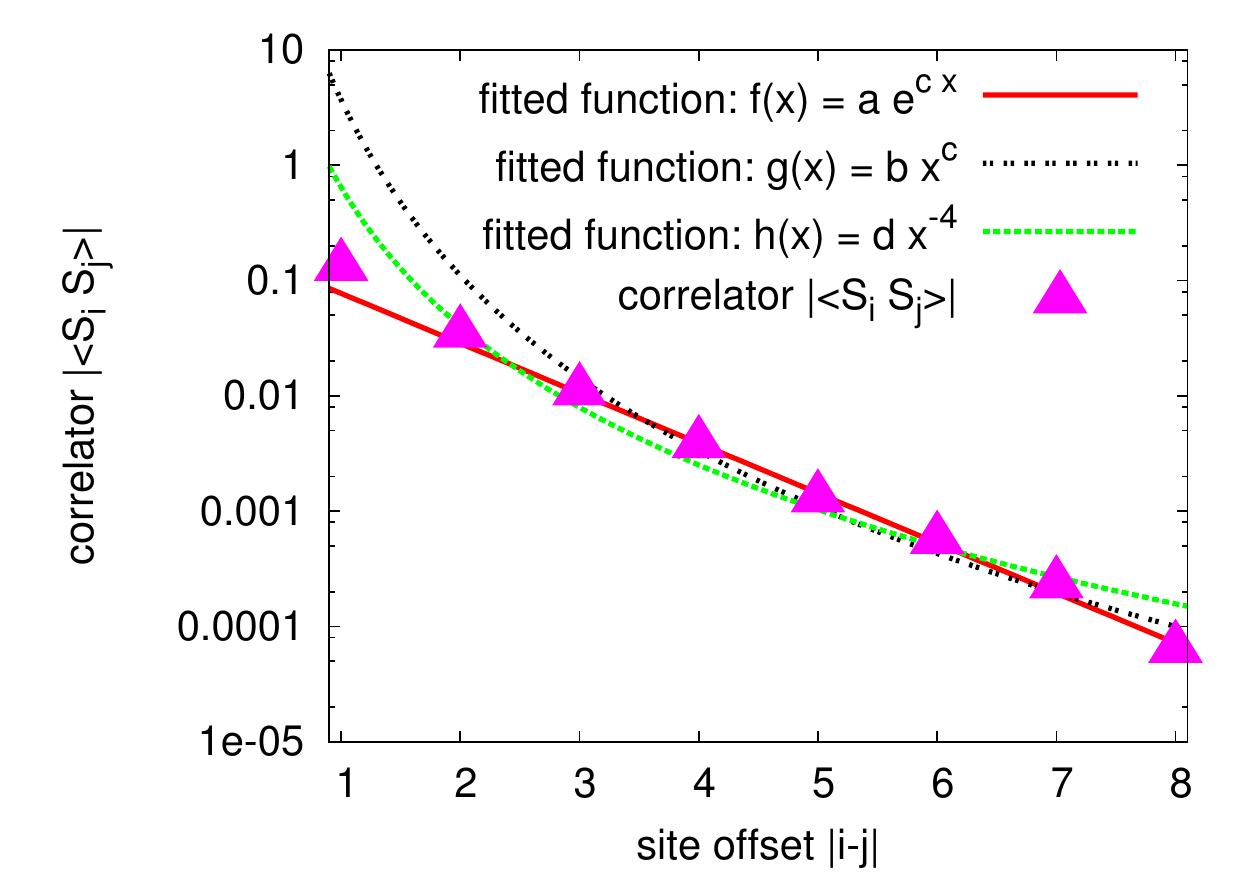}
}
% \hspace{2mm}
\subfigure[\label{fig:dimerfit}]{
\includegraphics[width=0.4\textwidth]{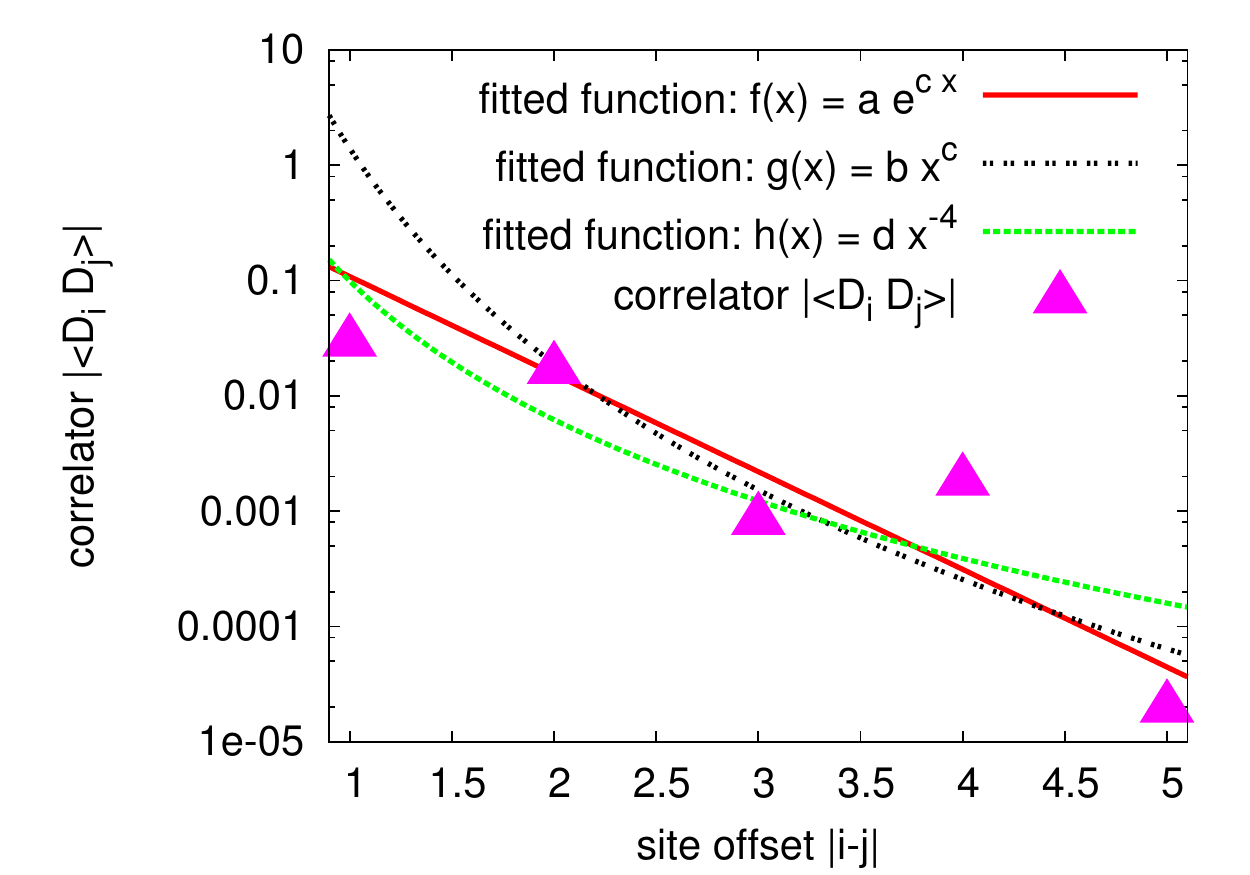}
}
\caption{\label{fig:correlators}Log-linear plots of the absolute value of the \ref{fig:corrfit} spin-spin and
\ref{fig:dimerfit} dimer-dimer correlation functions versus the distance $x=\vert i - j\vert$ for a XC12
(\ref{fig:corrfit}) and a YC8 (\ref{fig:dimerfit}) sample along one lattice axis with exponential and power law fits. An $x^{-4}$ line is shown as guide to the eye.}
\end{figure}

{\em Correlation Functions.}--For all cylinders, we find an antiferromagnetic spin-spin correlation function $\langle \vec{S}_i \cdot \vec{S}_j \rangle$ along different lattice axes with almost no directional dependence. Exponential fits with a very short correlation length of $\xi \simeq 1$ (Fig. \ref{fig:corrfit}) were consistently better than power law fits, in agreement with a spin gap. This is not consistent with an algebraic spin liquid \cite{hermele_properties_2008}, where the correlations are predicted to decay according to a power law $\sim \frac{1}{x^4}$.

\begin{figure}
\subfigure[\label{fig:XC16SF} XC16 system (196 sites)]{
   \includegraphics[width=110pt]{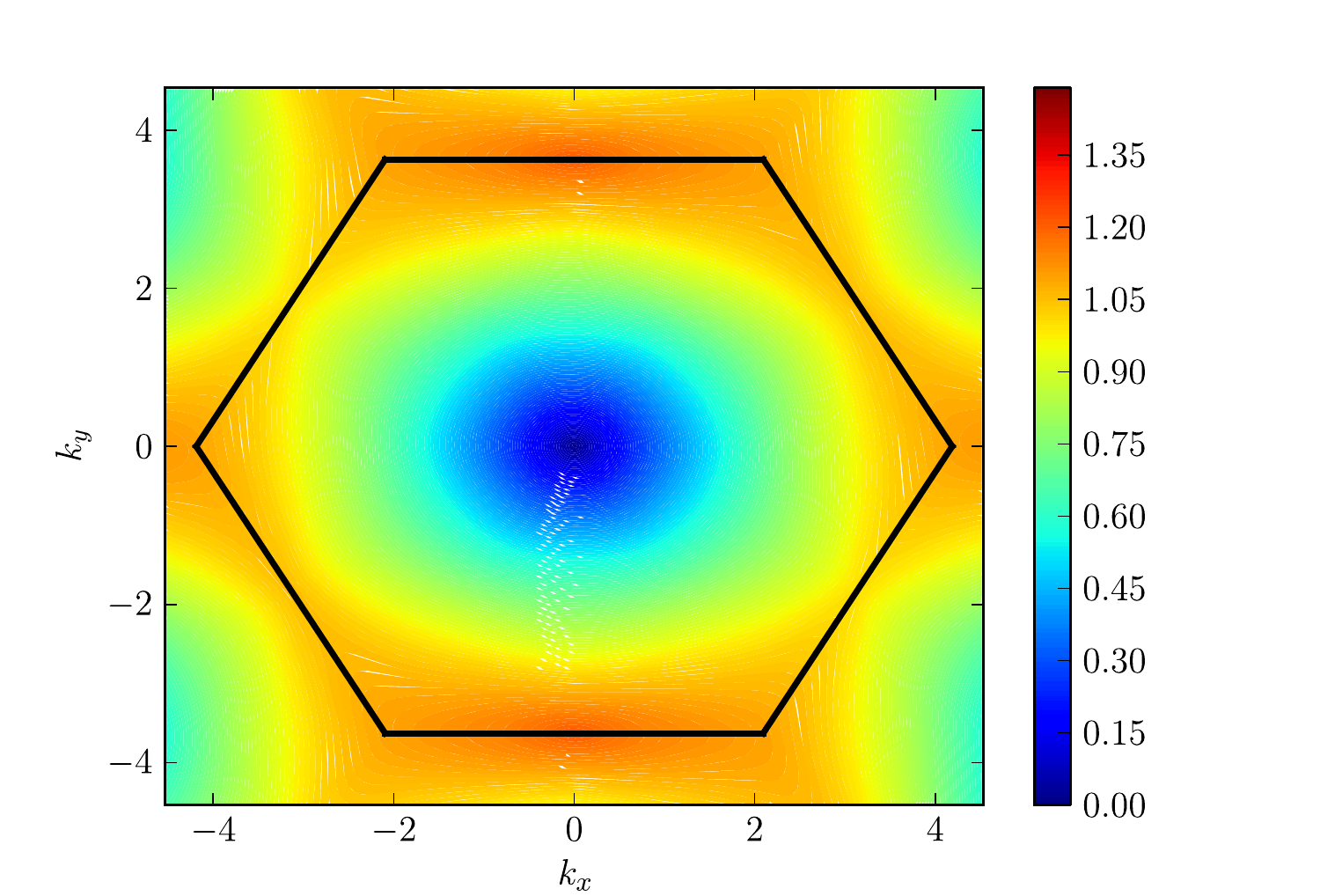}
   }
  \subfigure[\label{fig:KTSF} 27-site torus system]{
   \includegraphics[width=110pt]{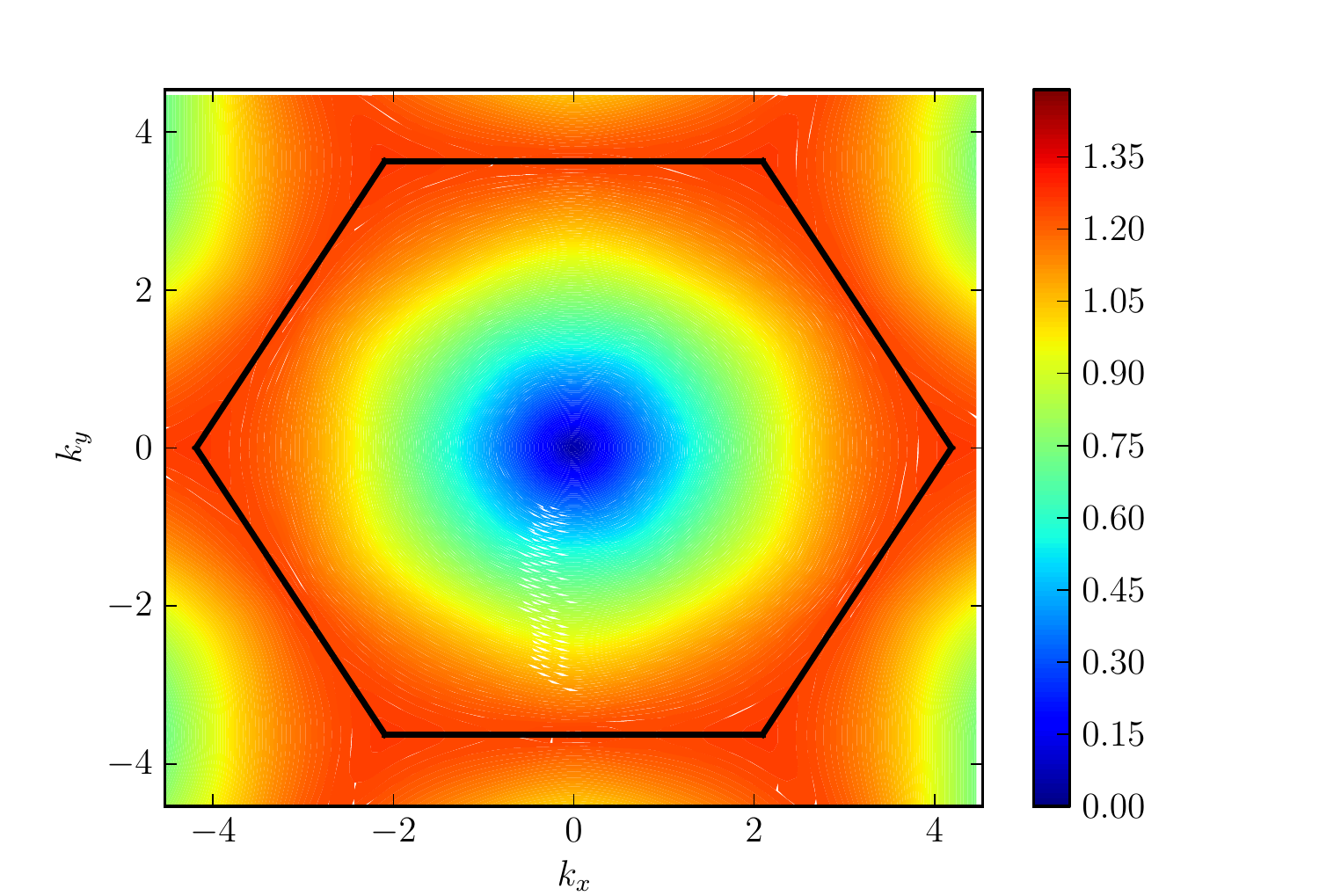}
   }
\caption{\label{fig:strucfactor}Two static structure factors $S(\vec{q})$; $k_x$, $k_y$ in units of reciprocal lattice basis vectors. Results are independent of the choice of 1D mapping (not shown).}
\end{figure}

We also consider the static spin structure factor $S(\vec{q}) = \frac{1}{N} \sum_{ij} e^{i \vec{q} \cdot (\vec{r}_i - \vec{r}_j)} \langle \vec{S}_i\cdot\vec{S}_j \rangle$, $\vec{q}$ in units of basis vectors $(\vec{b}_1,\vec{b}_2)$ of the reciprocal lattice. The spectral weight is concentrated evenly around the edge of the extended Brillouin zone, with not very pronounced maxima on the corners of the hexagon. Results for large cylinders agree well with ED results for tori up to 36 sites \cite{laeuchli_dynamical_2009}. All our $S(\vec{q})$ are in accordance with the prediction for a $\mathbb{Z}_2$ QSL\cite{sachdev_kagome-_1992}. 

We also find antiferromagnetically decaying, almost direction-independent dimer-dimer correlations for which again an exponential fit is favored (Fig.\ref{fig:dimerfit}), in agreement with a singlet gap.  Our data do not support the algebraic decay predicted \cite{hermele_properties_2008} for an algebraic QSL. 

\begin{figure}
  \includegraphics[height=160pt]{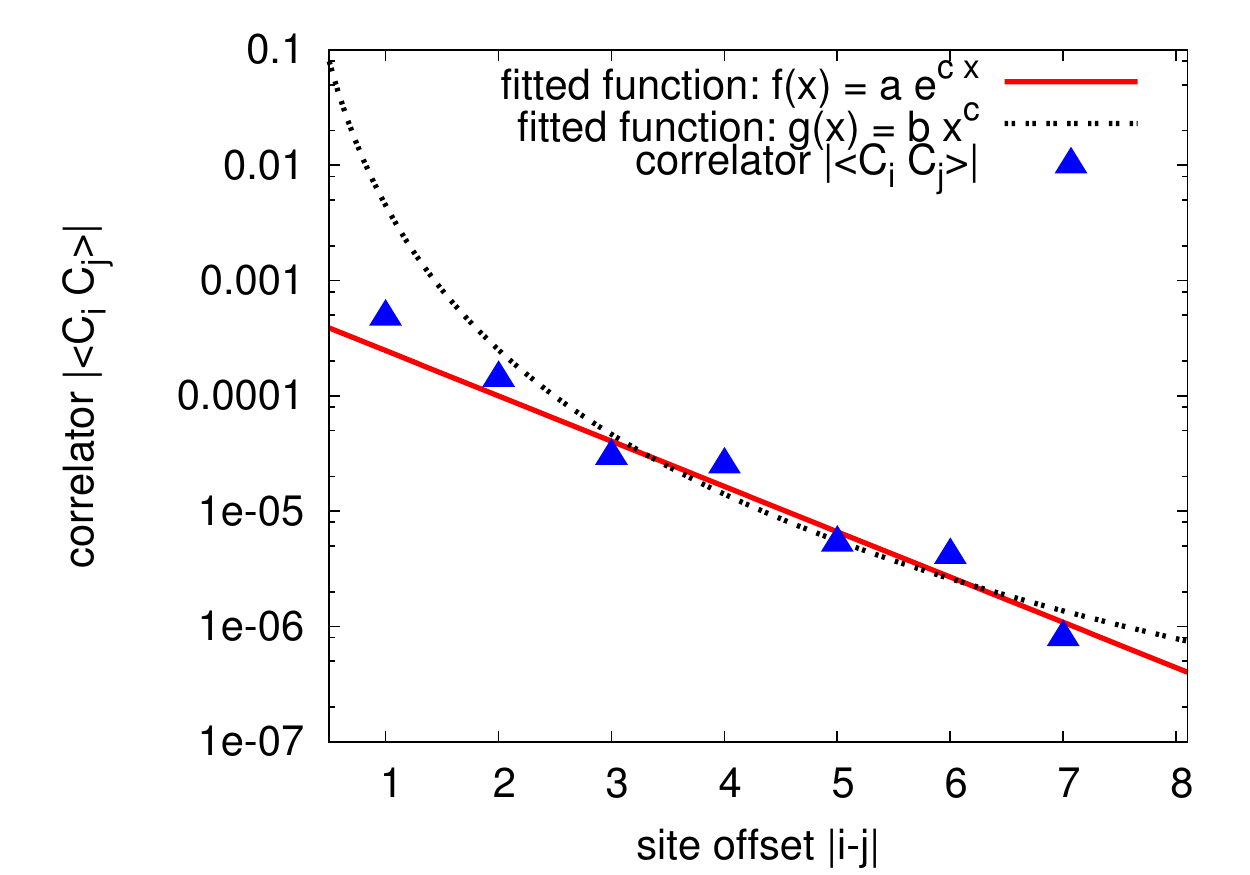}
  \caption{\label{fig:chiralfit}Log-linear plot of the absolute value of the chiral correlation function $\langle C_0, C_x \rangle = \langle \vec{S}_{i_0} \cdot (\vec{S}_{j_0}\times \vec{S}_{k_0}) \cdot \vec{S}_{i_x} \cdot (\vec{S}_{j_x}\times \vec{S}_{k_x})\rangle$ versus the distance $x = \vert \Delta_0 -\Delta_{x} \vert$ along a lattice axis for a 196-site YC8 sample with exponential and power law fits.}
\end{figure}
Chiral correlation functions \cite{waldtmann_chiral} $\langle C_{ijk} C_{lmn} \rangle = \langle \vec{S}_i\cdot(\vec{S}_j \times \vec{S}_k) \cdot \vec{S}_l \cdot(\vec{S}_m \times \vec{S}_n)\rangle$, where the loops considered are elementary triangles, did not show significant correlations for any distance or direction and decay exponentially (Fig. \ref{fig:chiralfit}), faster than the spin-spin correlations. Expectation values of single loop operators $C_{ijk}$ vanish, as expected for finite size lattices. Chiral correlators for other loop types and sizes decay even faster. Our findings do not support chiral spin liquid proposals \cite{messio_kagome_2011, Yang_QSL,wen_chiral}. 

{\em Topological Entanglement Entropy.}--To obtain direct evidence regarding a topological state, we consider the topological entanglement entropy \cite{Levin_topological_2006,kitaev_2006,balents_j1J2_2011}. For the ground states of gapped, short-ranged Hamiltonians in 2D, entanglement entropy scales as $\mathcal{S} \simeq c$, if we cut cylinders into two, with corrections in the case of topological ground states \cite{wen_mean-field_1991}. We examine Renyi entropies $S_{\alpha} = (1-\alpha)^{-1} \log_2 \textrm{tr} \rho^\alpha$, $0\leq\alpha<\infty$, where $\rho$ is a subsystem density matrix. Scaling is expected as $\mathcal{S}_{\alpha} \simeq \eta c - \gamma$ where $\eta$ is an $\alpha$-dependent constant. $\gamma$, the topological entanglement entropy, is independent of $\alpha$ \cite{zhang_topological_2011,flammia_topological_2009,zhang_entanglement_2011} and depends only on the total quantum dimension $D$ as $\gamma = \log_2(D)$ \cite{Levin_topological_2006,kitaev_2006}. In our mappings, DMRG gives direct access to density matrices of cylinder slices. We calculate $\mathcal{S}_{\alpha}$ for cylinders of fixed $c$ and extrapolate in $L^{-1}$ to $L\rightarrow\infty$;  a linear extrapolation in $c \rightarrow 0$ yields $\gamma$. Results are 1D mapping independent.
\begin{figure}
  \includegraphics[height=150pt]{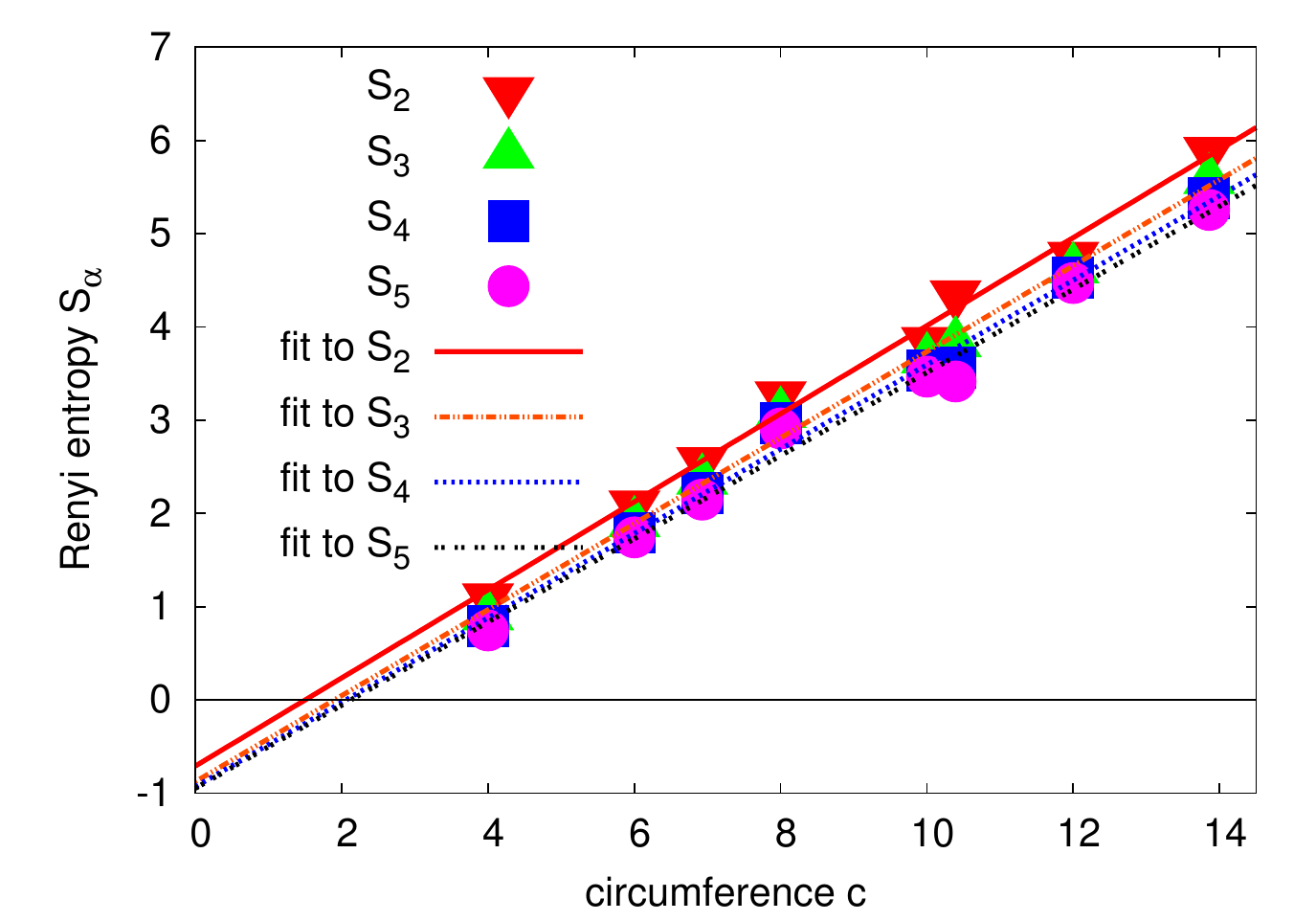}
  \caption{\label{fig:entent}Renyi entropies $\mathcal{S}_{\alpha}$ of infinitely long cylinders for various $\alpha$ versus circumference $c$,  extrapolated to $c=0$. The negative intercept  is the topological entanglement entropy $\gamma$.}
\end{figure}
We show intermediate values of $\alpha$ (Fig. \ref{fig:entent}), which all show a clearly finite value of $\gamma$, with a value very consistent with $\gamma=1$; large-$\alpha$ results agree. Small-$\alpha$ results are unreliable, as DMRG does not capture the tail of the spectrum of $\rho$ properly, but also point to a finite value of $\gamma$, hence a topological ground state. The quantum dimension is $D=2$, excluding chiral spin liquids ($\gamma=1/2$ or $D=\sqrt{2}$ \cite{zhang_topological_2011}). Rigorously, DMRG only provides a lower bound on $D$ \cite{zhang_grover_oshikawa_2011}, but the bound is essentially exact as DMRG is a method with low entanglement bias \cite{JiangBalents2012}.  

{\em Conclusion.}--Through a combination of a large number of DMRG states, large samples with small finite size effect, and the use of the SU(2) symmetry of the kagome model, we have been able to corroborate earlier evidence for a QSL as opposed to a VBC, due to energetic considerations and complete absence of breaking of space group invariance, although DMRG should be biased towards VBC due to its low-entanglement nature and the use of OBC. On the basis of the numerical evidence (spin gap, structure factor, spin, dimer and chiral correlations, topological entanglement entropy)
numerous QSL proposals can be ruled out for the kagome system. On the system sizes reached, the spin gap is very robust and essentially size-independent, ruling out all proposals for gapless spin liquids, consistent with the exponential decay of correlators. Individual gapless QSL proposals make other predictions not supported by numerical data, e.g. the static spin structure factor \cite{hermele_properties_2008}. Another strong observation is the very rapid decay of chiral correlations, ruling out proposals related to chiral QSL. The third strong observation is finite topological entanglement, which implies a topologically degenerate ground state for the kagome system. For quantum dimension 2, as found here, we have in principle, for a time-reversal invariant ground state, a choice between a $\mathbb{Z}_2$ phase and a double-semion phase \cite{levin_topological_2005,gu_2009}. A $\mathbb{Z}_2$ QSL emerges straightforwardly in effective field theories of the kagome model as a mean-field phase stable under quantum fluctuations breaking a U(1) gauge symmetry down to $\mathbb{Z}_2$ due to a Higgs mechanism \cite{Sachdev_08}, and microscopically a resonating valence bond state formed from nearest-neighbor Rokhsar-Kivelson dimer coverings of the kagome lattice directly leads to a $\mathbb{Z}_2$ QSL \cite{poilblanc_2012a,poilblanc_2012b} albeit for a variational energy far from the ground state energy. The concentration of weight of the structure factor at the hexagonal Brillouin zone edge with shallow maxima at the corners would also point to the $\mathbb{Z}_2$ QSL as proposed by Sachdev\cite{sachdev_kagome-_1992}, and a $\mathbb{Z}_2$ QSL is also consistent with all other numerical findings. All this provides strong evidence for the $\mathbb{Z}_2$ QSL, whereas to our knowledge, no plausible scenario for the emergence of a double-semion phase in the KAFM has been discovered so far, making it unplausible, but of course not impossible. An analysis of the degenerate ground state manifold as proposed in \cite{zhang_grover_oshikawa_2011}, not possible with our data, would settle the issue. Even if the answer provided final evidence for a $\mathbb{Z}_2$ QSL, many questions regarding the detailed microscopic structure of the ground state wave functions and the precise nature of the $\mathbb{Z}_2$ QSL would remain for future research.

S.D. and U.S. thank F. Essler, A. L\"auchli, C. Lhuillier, D. Poilblanc, S. Sachdev, R. Thomale and S. R. White for discussions. S.D. and U.S. acknowledge support by DFG. U.S. thanks the GGI, Florence, for its hospitality. I. P. M. acknowledges support from the Australian Research Council Centre of Excellence for Engineered Quantum Systems and the Discovery Projects funding scheme (Project No. DP1092513). 

{\em Note added.} --Recently, we became aware of  Ref.~\cite{JiangBalents2012} which calculates topological entanglement entropy from von Neumann entropy for a next-nearest neighbor modification of the KAFM, perfectly consistent with our results of $D=2$ for the KAFM itself.

\newpage

{\bf SUPPLEMENTARY INFORMATION}

\begin{figure}[h]
\subfigure[\label{fig:xpath}XC8 system, 38 sites]
  {
   \includegraphics[height=90pt]{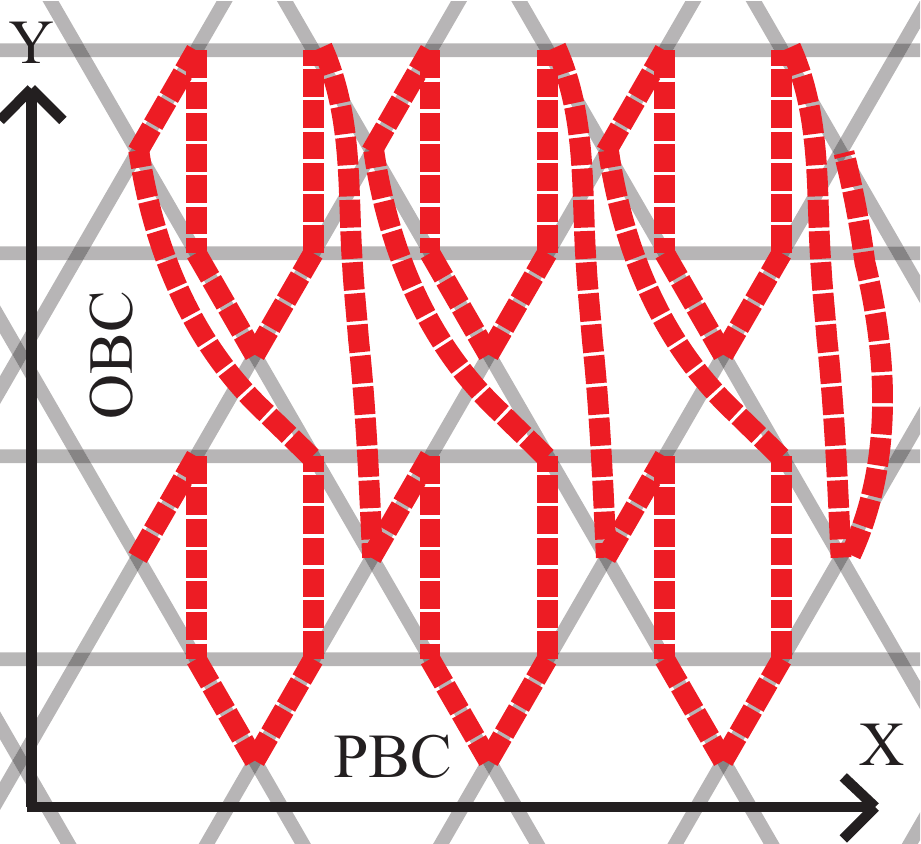}
   }
  \hspace{5mm}
  \subfigure[\label{fig:ypath} YC6 system, 39 sites]
  {
   \includegraphics[height=90pt]{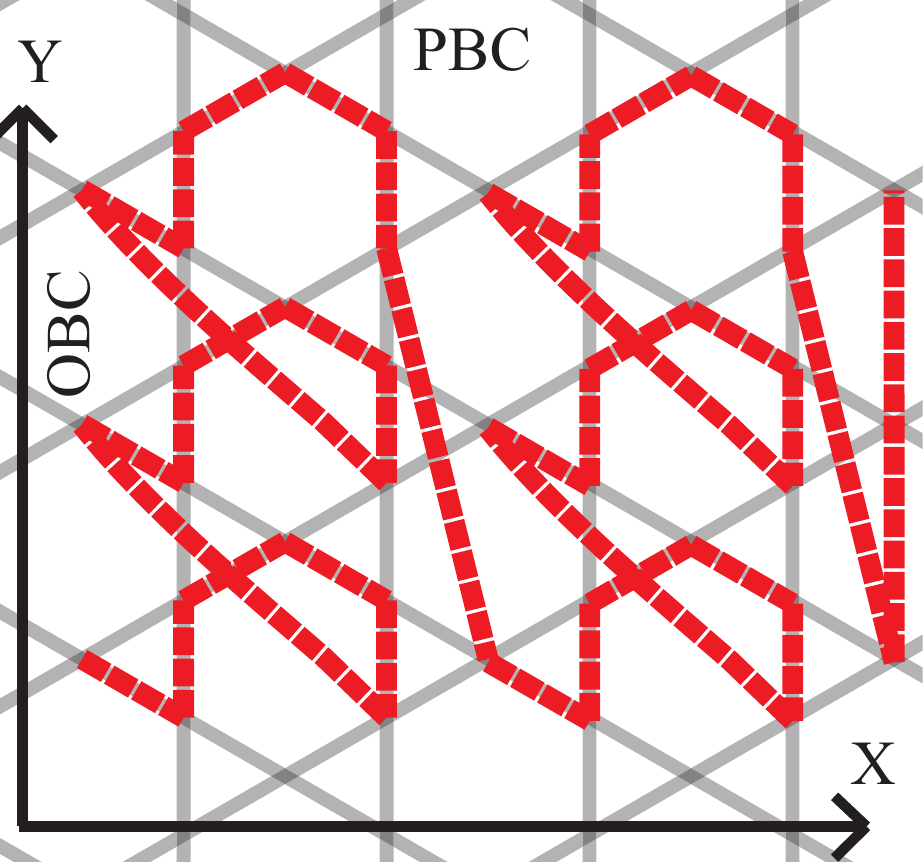}
   }
   \caption{\label{fig:path}(Color online) We map the two-dimensional system to a one-dimensional chain using two different mappings, one for aligning the lattice to the X-axis (a) and one for aligning it to the Y-axis (b) with periodic (open) boundary conditions in the vertical (horizontal) direction. The red broad line represents the one-dimensional chain.}
\end{figure} 

\begin{figure}
  \includegraphics[height=160pt]{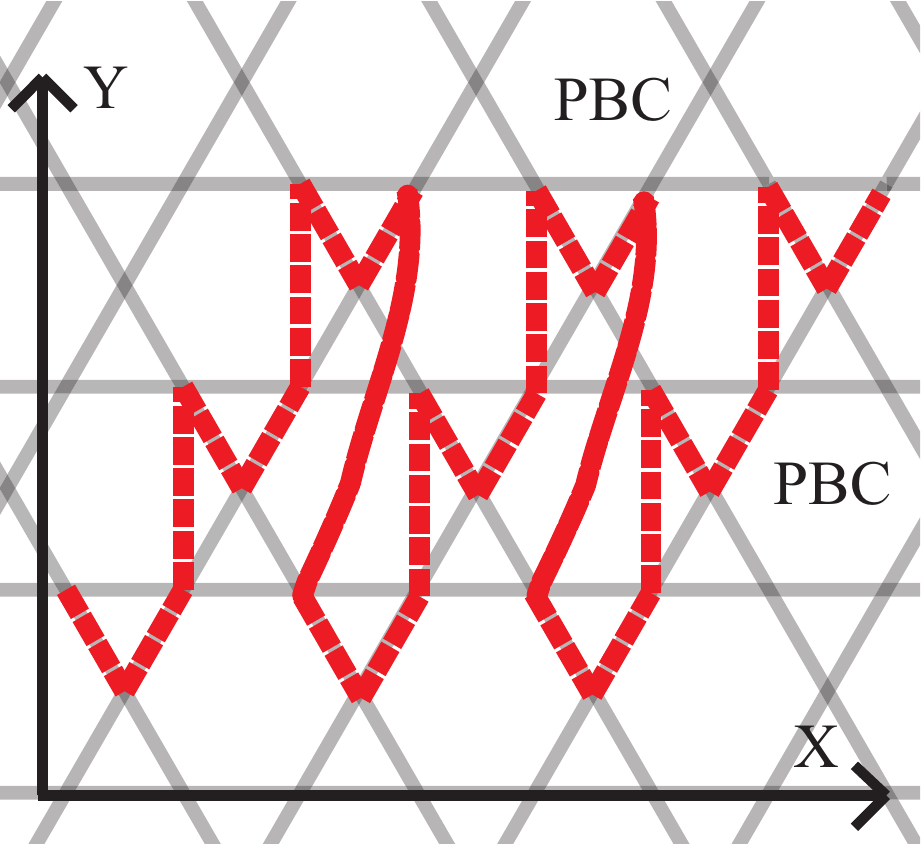}
  \caption{\label{fig:torus}(Color online) Path for a torus in DMRG}
\end{figure}

\begin{figure}[ht]
\subfigure[\label{fig:XC8GS} Ground state for $S = 0$.]{
   \includegraphics[width=165pt]{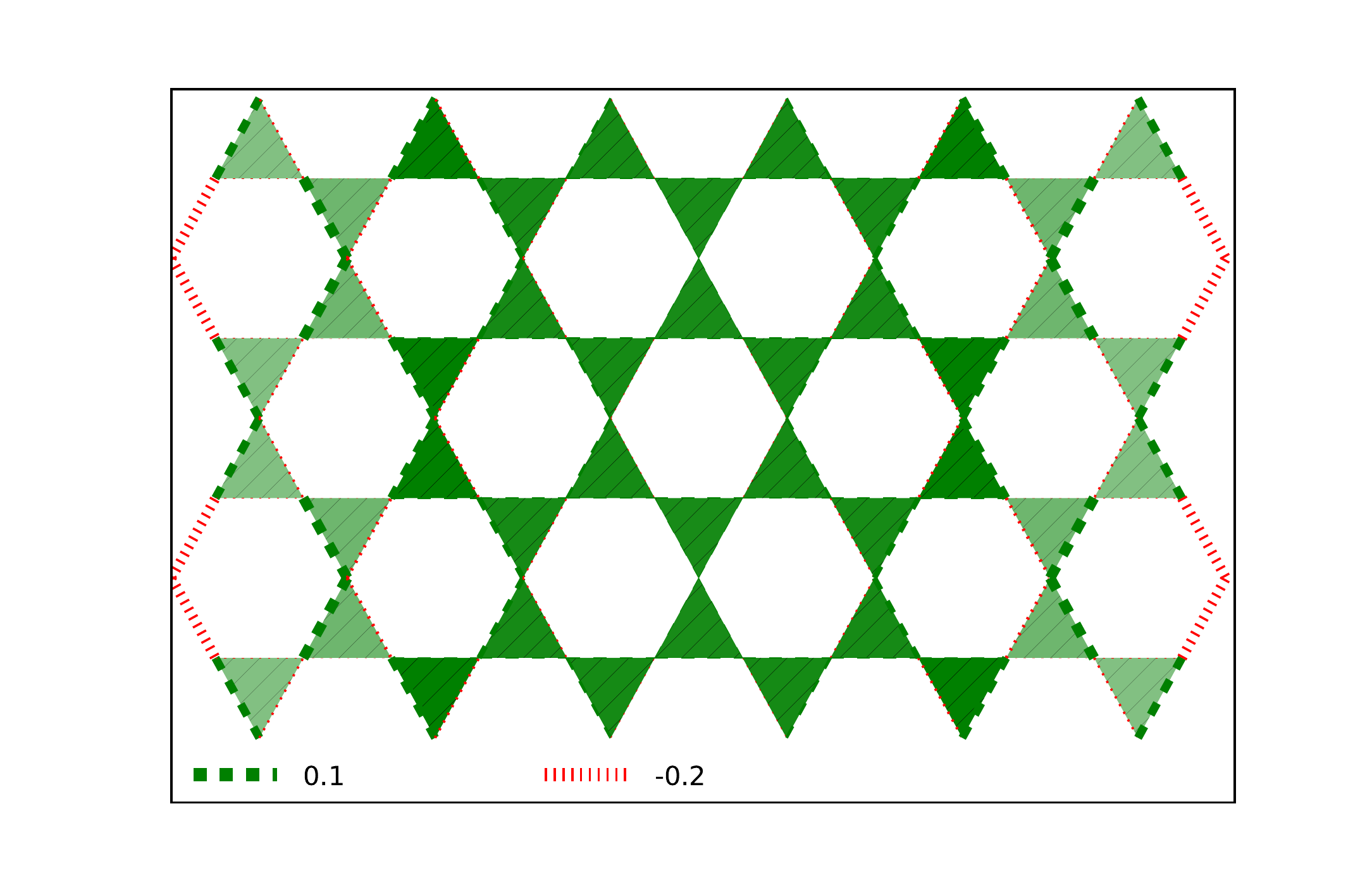}
   }
  \subfigure[\label{fig:XC8ES} Ground state for $S = 1$.]{
   \includegraphics[width=165pt]{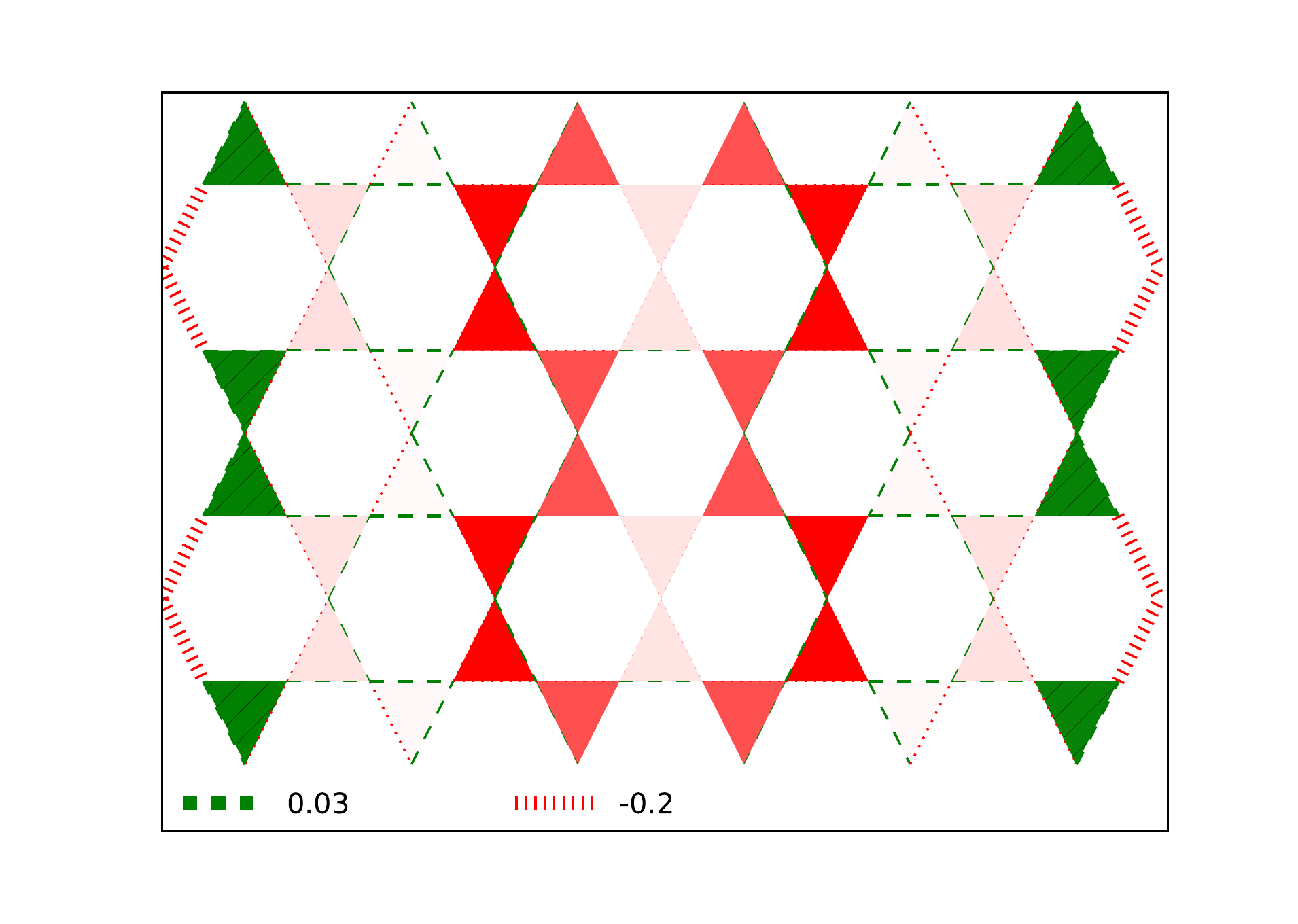}
   }
  \subfigure[\label{fig:XC8Diff} Difference in bond energies between the $S=0$ and the $S=1$ samples.]{
   \includegraphics[width=0.4\textwidth]{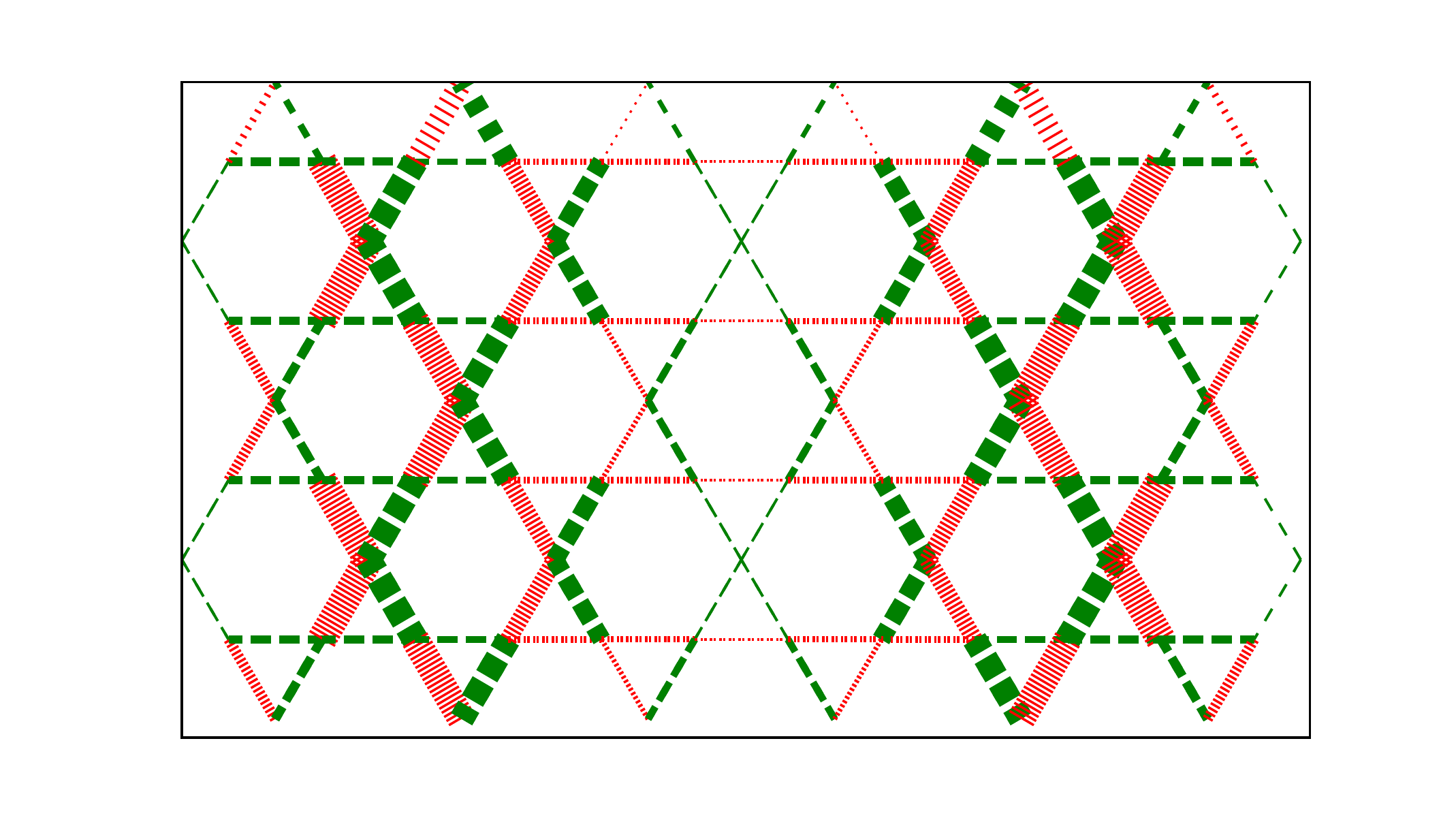}
   }
\caption{\label{fig:tricorr}(Color online) Bond and triangle energies for the ground state and lowest triplet excitation of a 74-site XC8 sample. In panels (a) and (b), the line width is proportional to $\vert \langle \vec{S}_i \cdot \vec{S}_j \rangle - e_0\vert$ where $e_0$ is the mean bulk energy. Green bonds denote $\langle \vec{S}_i \cdot \vec{S}_j \rangle < e_0$, red (dotted) bonds denote $\langle \vec{S}_i \cdot \vec{S}_j \rangle > e_0$. The triangle color (pattern) and intensity correspond to the deviation of the sum of the bond energies on the three triangle bonds from the mean $3e_0$, where the green (hatched) triangles denote a lower value, i.e. $\sum e_i < 3e_0$. In panel (c), bond energies of the lowest $S=0$ state are subtracted from those of the lowest $S=1$ state. The line width is proportional to the absolute value of the energy difference, green (hatched) lines correspond to positive and red (dotted) lines to negative energy differences.}
\end{figure}

\begin{figure}[ht]
\subfigure[\label{fig:vbc} This snapshot of a not yet converged calculation (insufficient number of DMRG states) shows bond energy patterns that break the translational invariance.]{
   \includegraphics[width=0.4\textwidth]{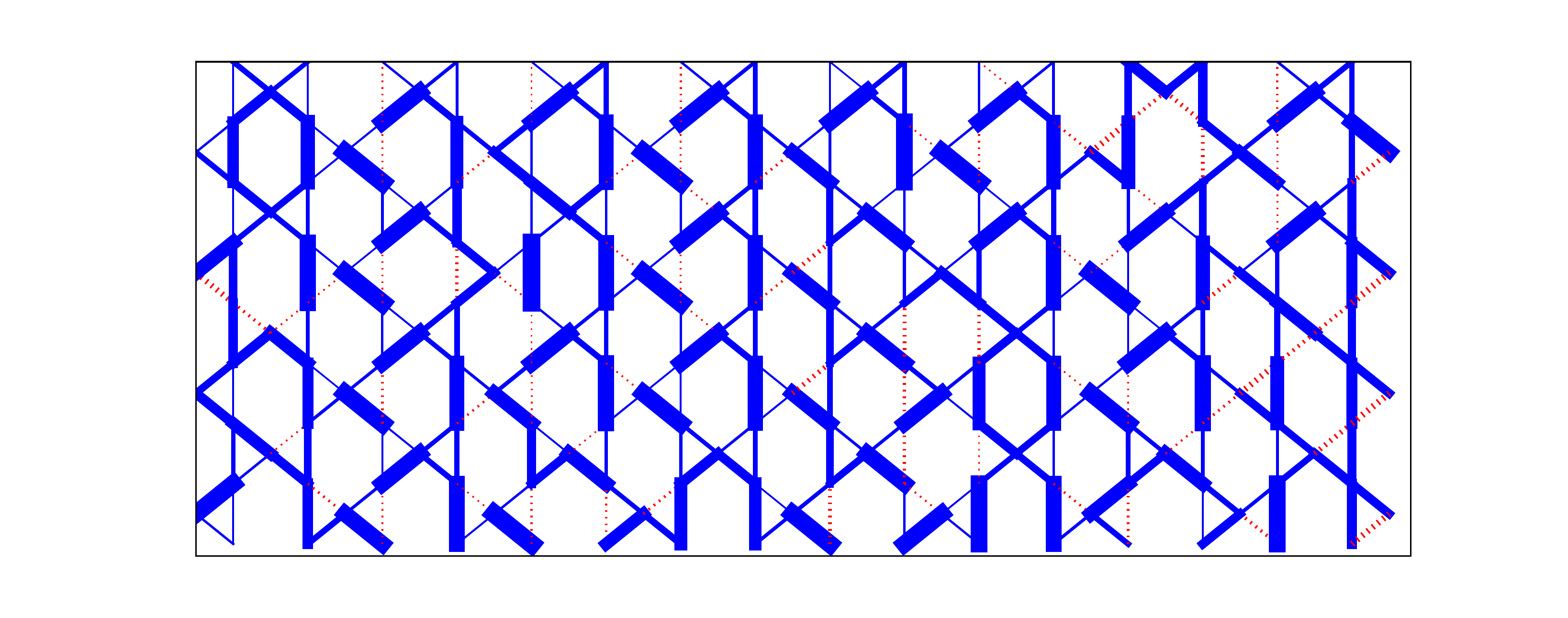}
   }
  \hspace{1mm}
  \subfigure[\label{fig:sl} This snapshot of a well converged calculation (sufficient number of DMRG states and sweeps) shows no pattern in the bond energies except for edge effects. In the bulk, a spin liquid state without breaking of translational invariance emerges.]{
   \includegraphics[width=0.4\textwidth]{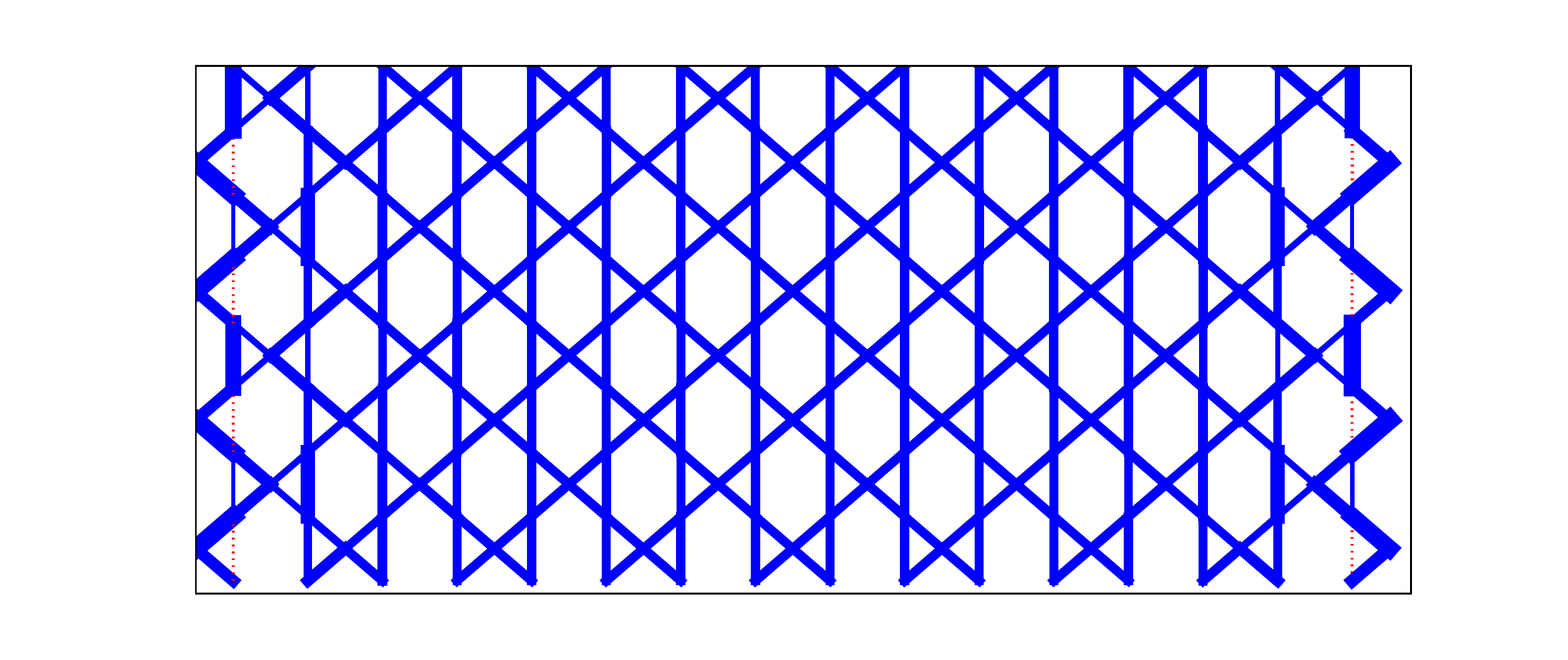}
   }
\caption{Visualization of the energy per bond for two snapshots in an iterative DMRG ground state calculation. The bond line width corresponds to the absolute value of the bond energy; the sign is negative (antiferromagnetic) for blue bonds, positive (ferromagnetic) for all red bonds, of which there are a few towards the edge.} 
\end{figure}

\begin{figure}
  \subfigure[\label{fig:YC8Hex}]{
    \includegraphics[width=115pt]{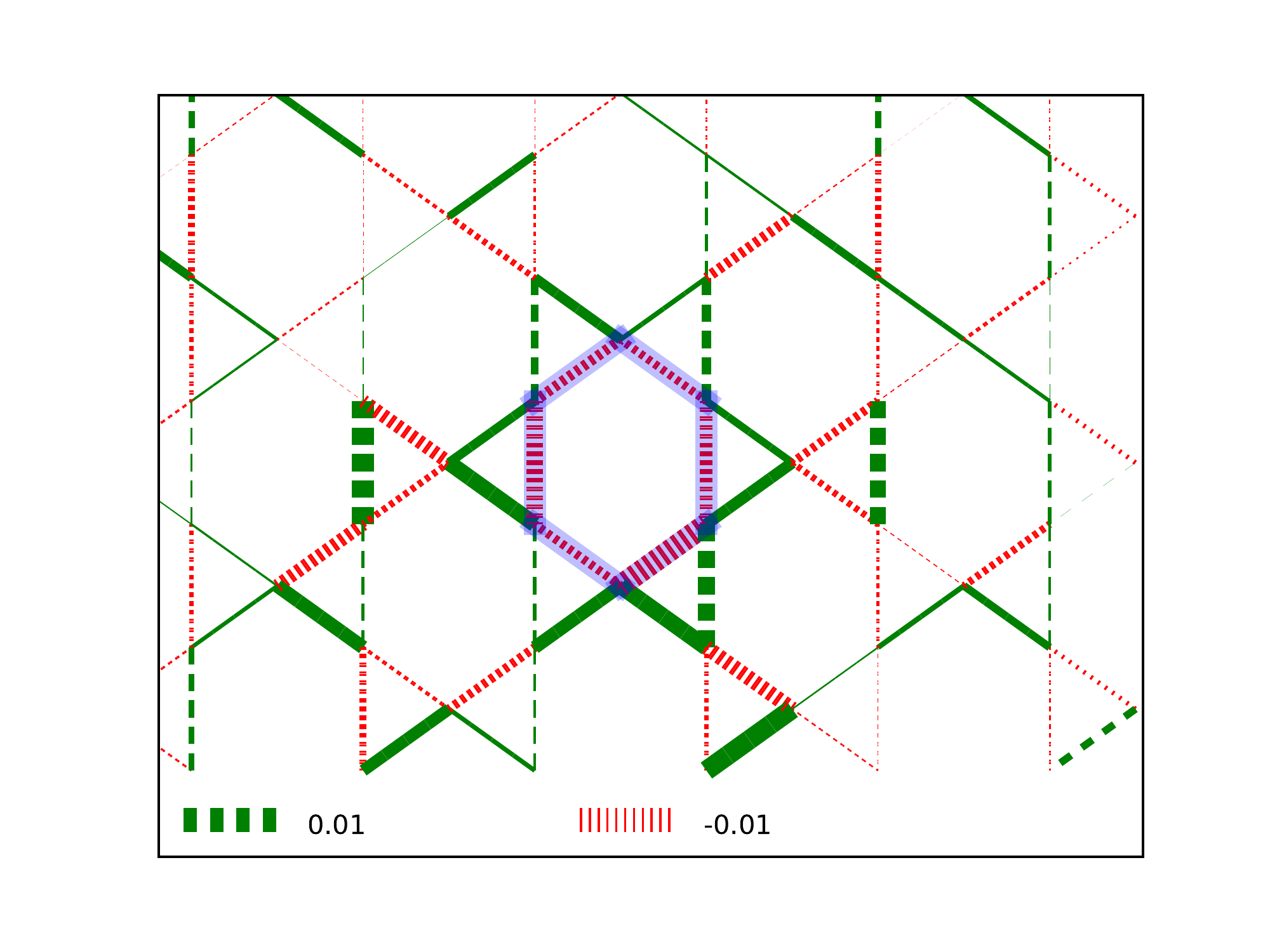}
  }
  \subfigure[\label{fig:YC8Diam}]{
    \includegraphics[width=115pt]{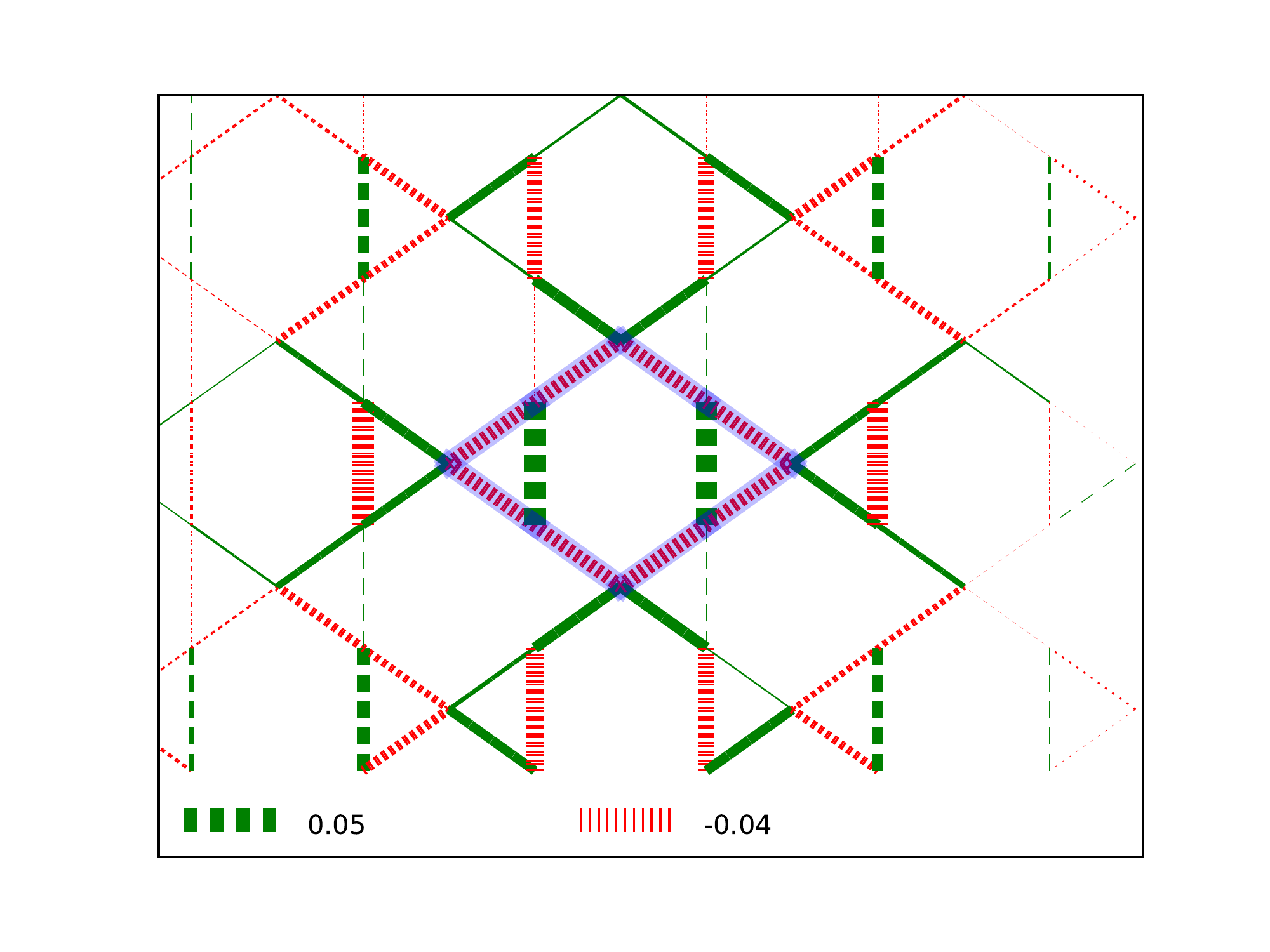}
  }
  \subfigure[\label{fig:YC8Vert}]{
    \includegraphics[width=0.4\textwidth]{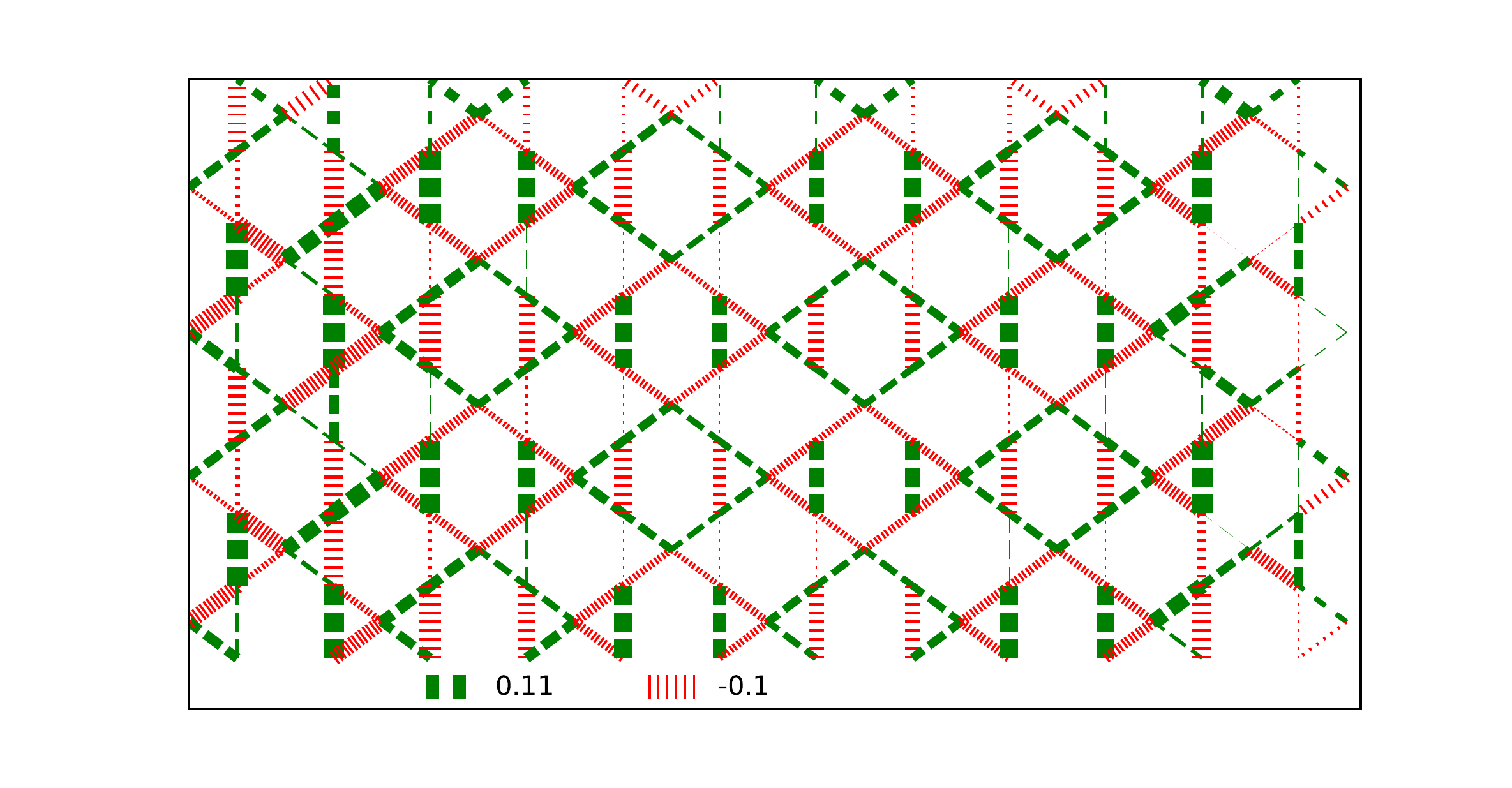}
  }
\caption{\label{fig:resonance}(Color online) Snapshots of the resonance pattern for a 150-site YC8 sample, using the same nomenclature as Fig. \ref{fig:tricorr}. Line widths correspond to the deviation of the bond energy from the mean bulk energy $e_0$; triangle color and intensity show the deviation of the sum of the three triangle bond energies from the bulk average $3e_0$. In each case, interactions on certain bonds (highlighted by color) have been enhanced, in \ref{fig:YC8Hex} a six-site hexagon,
in \ref{fig:YC8Diam} an eight-site diamond by $0.001$ each. In \ref{fig:YC8Vert}, the interaction strength of every second vertical bond was alternatingly changed by $\pm 0.5$\%, i.e. every fourth bond was strenghtened. The surrounding dimers arose in response to these changes, with the response increasing from \ref{fig:YC8Hex} to \ref{fig:YC8Diam} to \ref{fig:YC8Vert}.}
\end{figure}

\begin{figure}
    \includegraphics[width=0.5\textwidth]{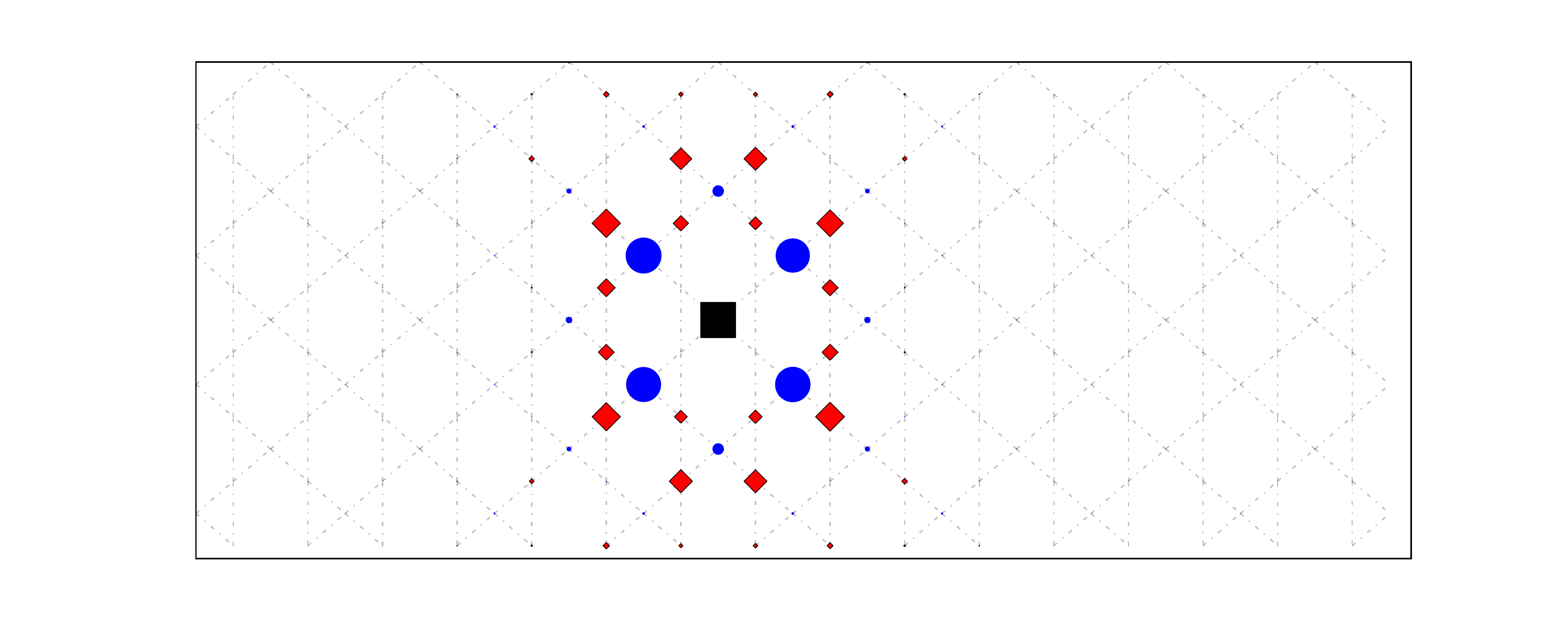}
  \caption{\label{fig:circlecorr} (Color online) Spin-spin correlations in the ground state of a 196-site YC8 sample system. The diameter of the circles (diamonds) is proportional to the absolute value of the spin-spin correlation with the central reference site (black square). Blue circles (red diamonds) denote positive (negative) correlations. Nearest-neighbor correlations have been left our for clarity. The lattice is drawn as a guide for the eyes.}
\end{figure}

{\em Mapping two-dimensional kagome lattices to one-dimensional chains for DMRG treatment.} As DMRG is a one-dimensional method, the two-dimensional kagome lattice on cylinders and tori has to be mapped to a one-dimensional chain with long-ranged interactions. There are multiple (in fact, combinatorially many) ways to map cylinders and tori to one-dimensional systems, however, ideally they keep interactions as short-ranged and the resulting path as regular as possible. Out of a large variety we tested we choose the two ways to map the kagome lattice to chains (Fig. \ref{fig:path}) that show the fastest convergence of energy in DMRG runs and label these either as X-cylinders (XC) or Y-cylinders (YC) depending on the lattice axes' alignment. In this notation, YC6 denotes a cylindrical system where one of the three lattice axes is aligned with the $y$-axis and a circumference $c$ of six lattice spacings. For XC systems (alignment of one of the lattice axes with the $x$-axis) the circumference is measured instead in units of $\sqrt{3}/2$ times the lattice spacing, so that e.g. the XC12 has a circumference of $c=6\sqrt{3}\approx 10.4$ lattice spacings. In the case of tori, which we considered mainly for reference purposes, only a single path was retained (Fig. \ref{fig:torus}). It is worthwhile to point out the path independence of results: where we consider the same cylinders as \cite{yan_spin_2011}, results do agree although they used yet another mapping.

{\em Identification of bulk vs.\ boundary excitations.} To rule out boundary excitations in the lowest $S=1$ state, we examine the difference in bond energies for the lowest lying states in the two spin sectors $S=0$ and $S=1$, finding no significant difference at the boundaries but a visible change in the bulk (Fig. \ref{fig:tricorr}).

{\em Supplementary information on ground state properties.} In order to exclude a valence bond crystal more rigorously, we consider the bond energies (nearest-neighbor correlators) where a valence bond crystal would exhibit a frozen pattern of different bond energies. We do not observe this for any of our ground states (see Fig. \ref{fig:tricorr}(a)). Interestingly, it turns out that we can see this frozen pattern in \textit{unconverged} wave functions (Fig. \ref{fig:vbc}). A further increase of the number of kept DMRG states and continued sweeping makes these patterns vanish in the bulk (Fig. \ref{fig:sl}). The presence of these frozen bond patterns hence is a distinguishing feature of an insufficiently converged wave function as it disappears upon lowering the energy and approaching the true ground state, where the bond energies only show deviations from the average at the cylinder's edges (Fig. \ref{fig:tricorr}). DMRG -- similar to other tensor network methods such as PEPS and MERA -- has a low-entanglement bias, because the underlying matrix product states structure can only capture entanglement up to a strength roughly logarithmic in the number of DMRG states: for an insufficient number of ansatz states, DMRG will therefore among states of similar energy prefer those of low entanglement, in our case valence bond crystals compared to quantum spin liquids.

To elicit additional information on the spin liquid state, we strengthen selectively the interaction on various patterns on some bonds, namely on a hexagon and on a diamond pattern and check whether this is taken up by the ground state structure (Fig. \ref{fig:resonance}(a) and (b)). In agreement with the U(1) DMRG calculation of \cite{yan_spin_2011}, we find in the SU(2) DMRG calculation that strengthening the interactions on the diamond pattern elicits the strongest response in the bond energies. Agreement is also obtained for modulating a pattern of every second vertical bond (Fig. \ref{fig:resonance} (c)), which finds an even stronger response; this was considered in \cite{yan_spin_2011} as evidence that the ground state of the kagome model arises from melting a valence bond state exhibiting a similar bond pattern.

As an additional check for preferred orderings, we also consider spin-spin correlations in real space (Fig. \ref{fig:circlecorr}) where lattice symmetry breaking orderings would show up as stronger correlations in certain directions. While we do not observe any signs for a valence bond crystal in the ground state, we see the band-like structure of the spin-spin correlations that was reported by L\"auchli \textit{et al.}\cite{lauchli_ground-state_2011} for tori. These pronounced staggered correlations along
selected loops wrapping around the sample are artifacts of the periodic boundary conditions and disappear for large circumferences. In agreement with expectations for a topological $\mathbb{Z}_2$ QSL we also observe the forming of band-like structures in the bond energies for cylinders with an odd number of sites (not shown).

\end{document}